\documentclass{aastex}

\usepackage{emulateapj5}

\usepackage{graphicx,xspace}
\newcommand\etal{{et al.}\xspace}
\newcommand\apec{{\sc apec}\xspace}
\newcommand\aped{{\sc aped}\xspace}
\newcommand\spex{{\sc spex}\xspace}
\newcommand\chandra{{\it Chandra}\xspace}
\newcommand\xmm{{\it XMM-Newton}\xspace}
\begin{document}

\title{Collisional Plasma Models with APEC/APED: Emission Line
Diagnostics of Hydrogen-like and Helium-like Ions}

\author{Randall K. Smith, Nancy S. Brickhouse} 
\affil{High Energy Astrophysics Division, Harvard-Smithsonian Center
for Astrophysics, \\ 
60 Garden Street, Cambridge, MA 02138; rsmith@hea-cfa.harvard.edu,
nbrickhouse@hea-cfa.harvard.edu} 

\author{Duane A. Liedahl}
\affil{Lawrence Livermore National Laboratory, Department of Physics
and Advanced Technologies\\ Lawrence Livermore National Laboratory \\
P.O. Box 808, L-41, Livermore, CA 94550; duane@leo.llnl.gov. }

\author{John C. Raymond} 
\affil{Solar and Space Physics Division, Harvard-Smithsonian Center
for Astrophysics, \\ 60 Garden St., Cambridge, MA 02138;
jraymond@cfa.harvard.edu}

\begin{abstract}
New X-ray observatories (\chandra and \xmm) are providing a wealth of
high-resolution X-ray spectra in which hydrogen- and helium-like ions
are usually strong features.  We present results from a new
collisional-radiative plasma code, the Astrophysical Plasma Emission
Code (\apec), which uses atomic data in the companion Astrophysical
Plasma Emission Database (\aped) to calculate spectral models for hot
plasmas.  \aped contains the requisite atomic data such as collisional
and radiative rates, recombination cross sections, dielectronic
recombination rates, and satellite line wavelengths.  We compare the
\apec results to other plasma codes for hydrogen- and helium-like
diagnostics, and test the sensitivity of our results to the number of
levels included in the models.  We find that dielectronic
recombination with hydrogen-like ions into high ($n=6-10$) principal
quantum numbers affects some helium-like line ratios from low-lying
($n=2$) transitions.
\end{abstract}

\received{\underline{                                 }}
\revised{\underline{                                  }}
\accepted{\underline{                                 }}

\section{Introduction}

Modeling emission from an optically-thin collisionally-ionized hot
plasma has been an on-going problem in astrophysics (Cox \& Tucker
1969; Cox \& Daltabuit 1971; Mewe 1972; Landini \& Monsignori Fossi
1972; Raymond \& Smith 1977; Brickhouse, Raymond \& Smith 1995).  Over
time, the codes used in these papers have been updated or completely
rewritten; current versions are the \spex code (Kaastra, Mewe \&
Nieuwenhuijzen 1996) and the {\sc chianti}\ code (Dere \etal 1997; Landi
\etal 1999).  The frequently used {\sc mekal}\ (Mewe-Kaastra-Liedahl) code
(Kaastra 1992, Liedahl, Osterheld, \& Goldstein 1995) embedded in
XSPEC (Arnaud 1996) uses data that are now in the \spex code; its
results are similar to \spex.  We describe here a new plasma emission
code \apec (Atomic Plasma Emission Code) along with the Atomic Plasma
Emission Database (\aped).

Our primary goal is to create plasma emission models that can be used
to analyze data from the high resolution X-ray spectrometers on the
\chandra and \xmm telescopes.  Due to limitations on computer speed,
memory size, and the available atomic data, early plasma codes
included only the strongest emission lines from each ion, and
``bundled'' nearby lines, reporting only their summed emission at a
single wavelength.  In addition, the data were stored in the code
itself, and could not be easily separated and studied.  In contrast,
we have strictly separated the atomic data from the code, storing the
atomic data in FITS format files, which collectively form \aped.  We
have endeavored to maintain the data in \aped as close as possible to
the original form, for simplicity and ease in error checking. In
addition, \aped contains error estimates for many wavelengths; errors
on other values in the database are in progress.

Currently, \aped contains data for well over a million lines, although
many of these are too weak to be observed individually.  An \apec
calculation done at low electron density (1 cm$^{-3}$), with cosmic
abundances (Anders \& Grevesse 1989) contains over 32,000 unique lines
whose peak emissivities at temperatures between $10^4-10^9$\,K exceed
$10^{-20}$\,photons cm$^3$\,s$^{-1}$.  Although the literature of
theoretical calculations for observable atomic transitions is far from
complete, multiple calculations for some important rates do exist.
\aped contains all the different datasets we have collected ({\it
e.g.}\ the entire {\sc chianti}\ v2.0 database (Landi \etal 1999), as
well as the data referred to in this paper) to allow easy comparison
of different rates, and of the effect of different assumptions on the
emissivities of selected lines.  In a subsequent paper, we plan to
present a complete overview of the data in \aped and the emissivity
tables over the X-ray range.  Here we discuss what effect changing the
underlying atomic data has on hydrogen-like (H-like) and helium-like
(He-like) oxygen ions.

\section{Method}

\apec calculates both line and continuum emissivities for a hot,
optically-thin plasma which is in collisional ionization equilibrium.
Although \apec can calculate the ionization balance directly (and thus
handle non-equilibrium conditions), \aped does not yet contain all the
necessary ionization/recombination rates, so we use tabulated values
for the ionization balance in thermal collisional equilibrium.  We
primarily use the ionization balance calculated by Mazzotta \etal
(1998; hereafter MM98) because it is a recent compilation that
self-consistently treats all the astrophysically relevant ions.  \aped
itself contains recombination (radiative and dielectronic) rate
coefficients for oxygen.  We combined these with the ionization rates
from MM98) to create our own self-consistent ionization balance for
these ions and found some small differences ($< 25$\%; see
Figure~\ref{fig:LineRatio}(b)).  We do plan to include in \apec/\aped
a self-consistent calculation of the ionization balance and level
population in the future.

Since the term ``emissivity'' has a number of definitions, we give our
definition explicitly.  Similar to the terminology of Raymond \& Smith
(1977) (but in photon instead of energy units), the emissivity of a
spectral line is the total number of radiative transitions per unit
volume, divided by the product of the electron density $n_e$\ and the
hydrogen (neutrals and protons) density $n_H$ in the astrophysical
plasma.  The line emissivity therefore has units of photons
cm$^3$\,s$^{-1}$.  Since the number of photons emitted is actually
proportional to the density of the ions involved, this definition
implicitly requires an elemental abundance and ionization balance for
the relevant ion to be specified.

For a given electron temperature $T$\ and electron density $n_e$, the
level populations for each ion are calculated from the collisional
(de-)excitation rate coefficients, the radiative transition rates, and
the radiative and dielectronic recombination rate coefficients.
Excitation-autoionization processes have not yet been included in
\aped, but these are not significant for the H- and He-like ions in
the equilibrium plasmas discussed here.  For the H- and He-like
isosequences, we have data for all singly-excited levels up to
principal quantum number $n=5$\ with the exception of \ion{O}{7},
where we have data up to $n=10$.

For the He-like ion \ion{O}{7}, the collisional excitation rate
coefficients are from Sampson, Goett \& Clark (1983), Kato \& Nakazaki
(1989), and Zhang \& Sampson (1987) for the levels up to $n=5$, and
from HULLAC (Bar-Shalom, Klapisch, \& Oreg 1988; Klapisch \etal 1977)
calculations for $n=6-10$.  The radiative transition rates for $n \le
5$\ are from TOPbase (Fernley, Taylor \& Seaton 1987), Derevianko \&
Johnson (1997), Lin, Johnson, \& Dalgarno (1977), and from the NIST
database\footnote{http://physics.nist.gov/cgi-bin/AtData/main\_asd};
for $n > 5$, the rates are again from HULLAC.  The wavelengths are
taken from Drake (1988).  For the H-like ion \ion{O}{8}, we use the
collisional rate coefficients from Sampson, Goett \& Clark (1983).  We
have also compared with scaled values from Kisielius, Berrington \&
Norrington (1996), but find few significant differences in the
low-temperature region where the data overlap.  The radiative
transition rates are again from TOPbase and Shapiro \& Breit (1959)
for the two-photon ($1s2s {}^1$S$_0\rightarrow1s^2\,{}^1$S$_0$) rate.
The wavelengths are from Ericsson (1977).  For both ions, we use the
dielectronic recombination (DR) rate coefficients and satellite line
wavelengths from Vainstein \& Safronova (1978, 1980) for $n \le 3$.
For $n=4,5$\ we used data from Safronova, Vasilyev \& Smith
(2001). Dielectronic recombination of \ion{O}{8} into levels $n=6-10$\
of \ion{O}{7} are from Safronova (2001, private communication).
Radiative recombination is calculated using the Milne relation and the
photoionization cross-sections of Verner \& Yakovlev (1995) for the
ground states and Clark, Cowan, \& Bobrowicz (1986) for the
recombination to excited states.

\section{Results}

\begin{figure*}[t]
\includegraphics[totalheight=2.2in]{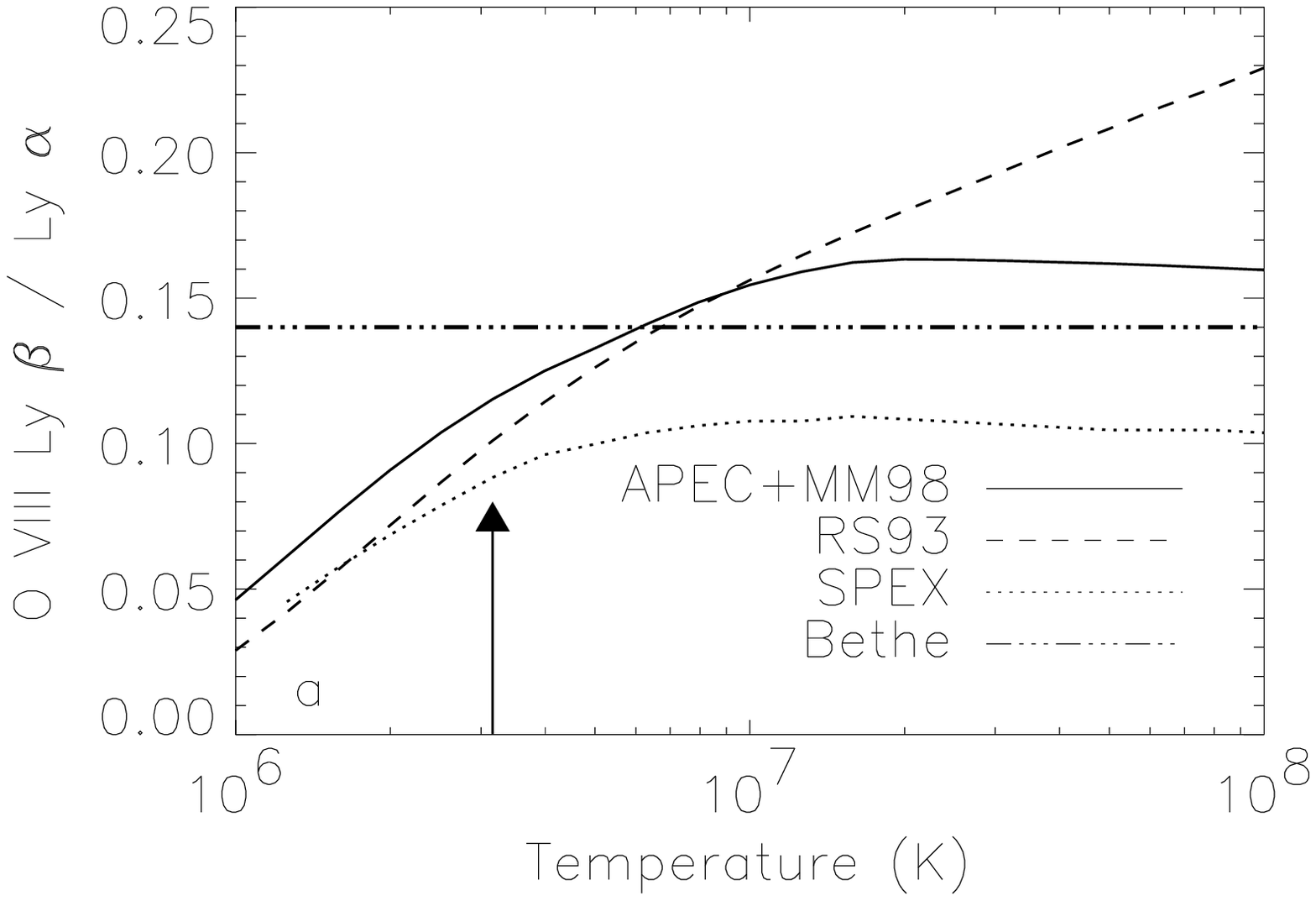}
\includegraphics[totalheight=2.2in]{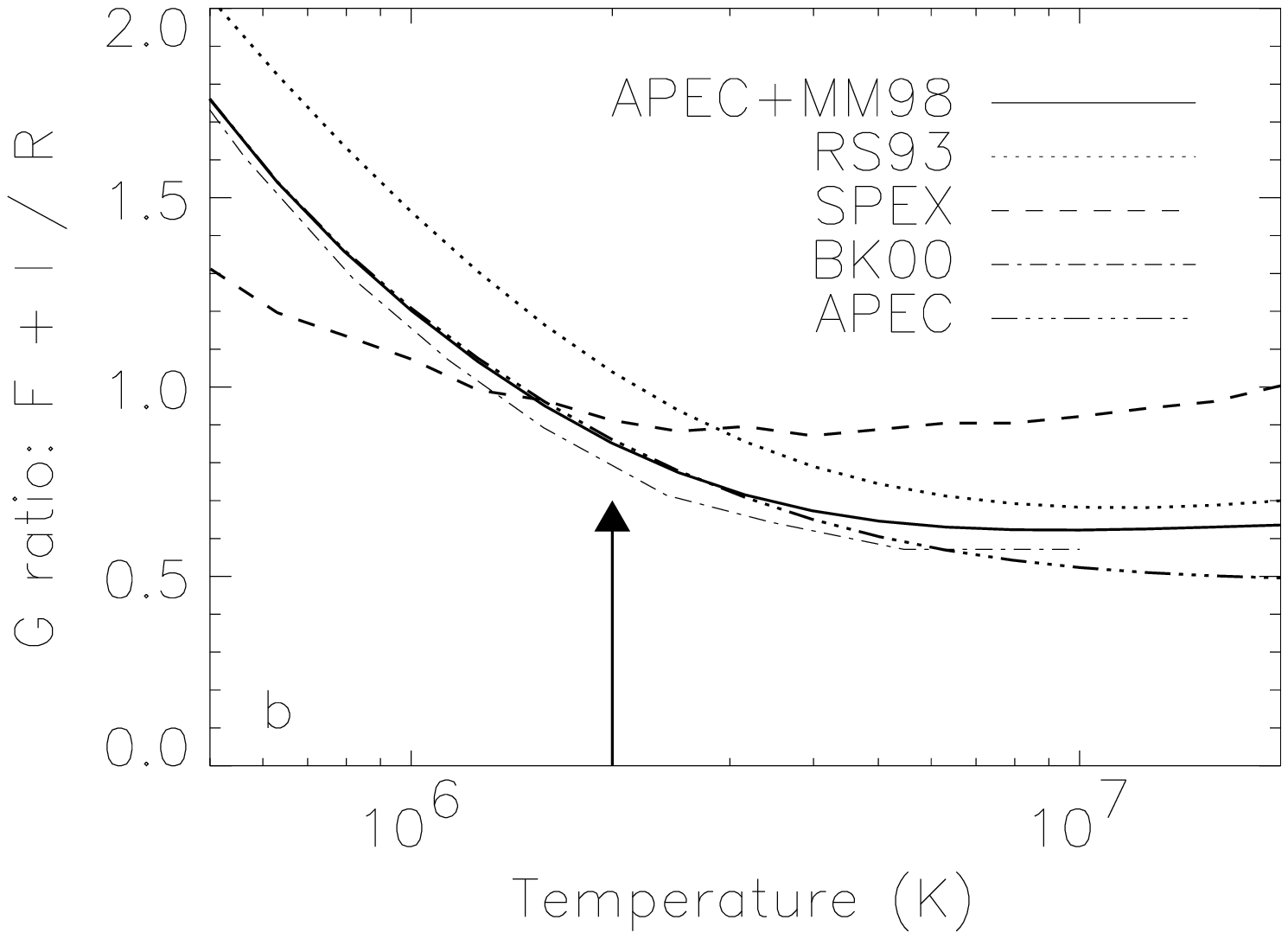}
\caption{(a) \protect{\ion{O}{8}}\ Ly$\beta$/Ly$\alpha$\ ratio [photon
units] as a function of temperature for a low electron density plasma
in collisional ionization equilibrium.  At the temperatures plotted,
each line emissivity is $\gtrsim 10^{-18}$\,ph cm$^3$s$^{-1}$, or
about 1/500th of the peak emissivity. The dot-dashed horizontal line
marks the Bethe ratio.  The arrow at $3.16\times10^6$\,K marks the
temperature of maximum total line emission.  (b) The
\protect{\ion{O}{7}} G ratio [photon units], as a function of electron
temperature for the same plasma codes plus data from Bautista \&
Kallman (2000) and a pure-\aped calculation, including the ionization
balance.  For the temperature range shown, the F+I and R line
emissivities exceed $10^{-18}$\,ph cm$^3$s$^{-1}$. The arrow at
$2\times10^6$\,K again marks the temperature of maximum total line
emission.\label{fig:LineRatio}}
\end{figure*}

We choose these oxygen ions for our comparisons since they are
relatively simple and have strong emission lines in many astrophysical
plasmas.  We will concentrate on two line ratios: for H-like ions,
Ly$\beta$(16.020\AA)/Ly$\alpha$(18.987\AA), and for He-like ions, the
frequently used $G \equiv (F+I)/R$\ ratio, where $F$\ (22.098\AA) is
the forbidden transition $1s2s\ {}^3\hbox{S}_1\rightarrow 1s^2\
{}^1\hbox{S}_0$, $I$\ (21.804\AA, 21.801\AA) is the sum of the two
intercombination transitions $1s2p\ {}^3\hbox{P}_{1,2}\rightarrow
1s^2\ {}^1\hbox{S}_0$, and $R$\ (21.602\AA) is the resonance
transition $1s2p\ {}^1\hbox{P}_1\rightarrow 1s^2\ {}^1\hbox{S}_0$.
Figure~\ref{fig:LineRatio}(a) shows the H-like Ly$\beta$/Ly$\alpha$\
line ratio from \apec (with the MM98 ionization balance), \spex 1.1,
and Raymond-Smith.  The {\sc mekal}\ results track those of \spex 1.1, and so
are not independently considered.  We also did a ``pure-\apec''
calculation where the ionization balance was found using recombination
rate coefficients taken from \aped.  However, this did not affect the
\ion{O}{8} Ly$\beta$/Ly$\alpha$\ ratio and so it is not shown.
Figure~\ref{fig:LineRatio}(b) shows the He-like G ratio from the same
calculations plus those of Bautista \& Kallman (2000) and the
pure-\apec calculation.  The difference between the \apec+MM98 and the
pure-\apec calculation for \ion{O}{7}, and the lack of a change in
\ion{O}{8}, can be understood in terms of the excitation mechanisms
for the lines, as will be discussed below.

Figure~\ref{fig:LineRatio}\ demonstrates two important points.  First,
near the peak emissivity of the emission lines, the code predictions
agree to within 25\%.  Second, substantial disagreements (exceeding
50\%) exist even for these simple ions when they are not near peak
emissivity.  Of course, comparing three theoretical models against
each other does not provide an accurate estimate of the model error.
For one limiting case, we can compare with analytic results.  In
Figure~\ref{fig:LineRatio}(a), the Bethe limit of 0.14 [photon units]
(Burgess \& Tully 1978) is also plotted; this is the ratio of the
electron collisional excitation rates in the high temperature limit,
taking into account the branching ratio of Ly$\beta$\ from the $3p$\
level.  In the H-like case, this limit should be a good estimate of
the Ly$\beta$/Ly$\alpha$\ ratio since collisional excitation is the
dominant method of populating excited levels.  We calculate that
cascades from collisional excitation to higher levels as well as
radiative recombination contribute only $\sim 10$\% of the direct rate
to each line, somewhat more for Ly$\beta$\ than Ly$\alpha$.  Since
dielectronic recombination cannot occur onto bare ions, it is not an
issue.  The agreement between the \apec result and the Bethe limit is
thus reassuring; as expected, \apec is slightly larger due to cascades
and radiative recombination.  This also explains why changing the
ionization balance (the pure-\apec case in
Figure~\ref{fig:LineRatio}(a)) does not affect the \ion{O}{8}
Ly$\beta$/Ly$\alpha$\ ratio.  Both lines are primarily populated by
direct collisions of electrons with \ion{O}{8}, so small changes in
the \ion{O}{8} abundance cancel.

Some \chandra and \xmm measurements of the \ion{O}{8}
Ly$\beta$/Ly$\alpha$\ ratio from astrophysical plasmas already exist.
The \chandra HETG observation of the RS CVn star II~Peg obtained a
value of $0.201 \pm 0.015$\ for this ratio (Huenemoerder 2000, private
communication).  In addition, both \chandra HETG and \xmm RGS
observations of the oxygen-rich SNR E0102 have been made, and although
the extended nature of the source makes the analysis more difficult,
the \ion{O}{8} Ly$\beta$/Ly$\alpha$\ ratio is measurable.  The
\chandra HETG obtained $0.138 \pm 0.016$ (Davis \etal 2001), while the
\xmm RGS1 and RGS2 instruments obtained $0.225 \pm 0.02$\ (Rasmussen
\etal 2001).  The discrepancy between these results is not yet
understood.  In addition, some laboratory measurements with the LLNL
EBIT of oxygen Ly$\beta$/Ly$\alpha$\ have been made.  They find a
range of values from 0.16 to 0.24 for electron energies above 1 keV
(Gendreau 2001, private communication).

Although the Bethe limit of 0.14 for the Ly$\beta$/Ly$\alpha$\ ratio
is a good estimate if collisional excitation dominates the \ion{O}{8}
excitation, other effects could change the observed value.  Line
blending, especially of \ion{Fe}{18}\ and \ion{Fe}{19}\ lines with O
Ly$\beta$, can increase the apparent Ly$\beta$\ emission.  Pure
radiative recombination also leads to a larger value of
Ly$\beta$/Ly$\alpha$ in H-like ions, similarly to the well-known
effects on line ratios for He-like ions in photoionized plasmas
(Porquet \& Dubau 2000; Bautista \& Kallman 2000).  However, tests
with \apec show that a plasma has to be nearly pure O$^{8+}$\ for this
to occur in a hot ($T > 5\times10^6$\,K) plasma.  Charge exchange into
excited levels has been suggested by Rasmussen \etal (2001) as a
possible way of increasing the ratio in E0102, but this requires
mixing of neutral material directly into highly ionized regions;
furthermore, it is difficult to imagine how this could be occurring in
II~Peg.  Finally, resonance scattering could also cause a change in
the observed ratio, with the size and direction of the effect
dependent on the spatial configuration of the plasma (e.g. Wood \&
Raymond 2000).

\begin{figure*}
\includegraphics[totalheight=2.2in]{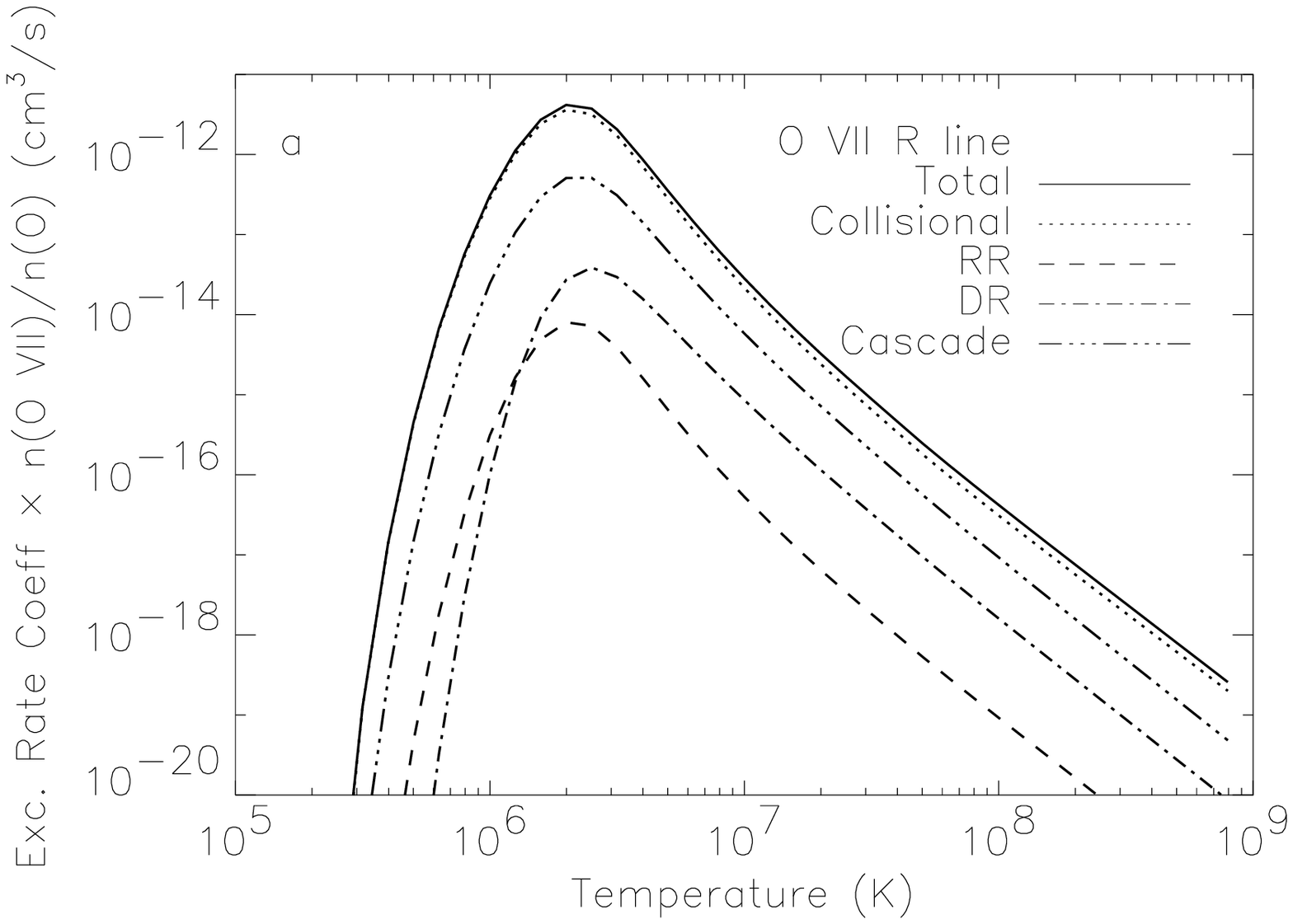}
\includegraphics[totalheight=2.2in]{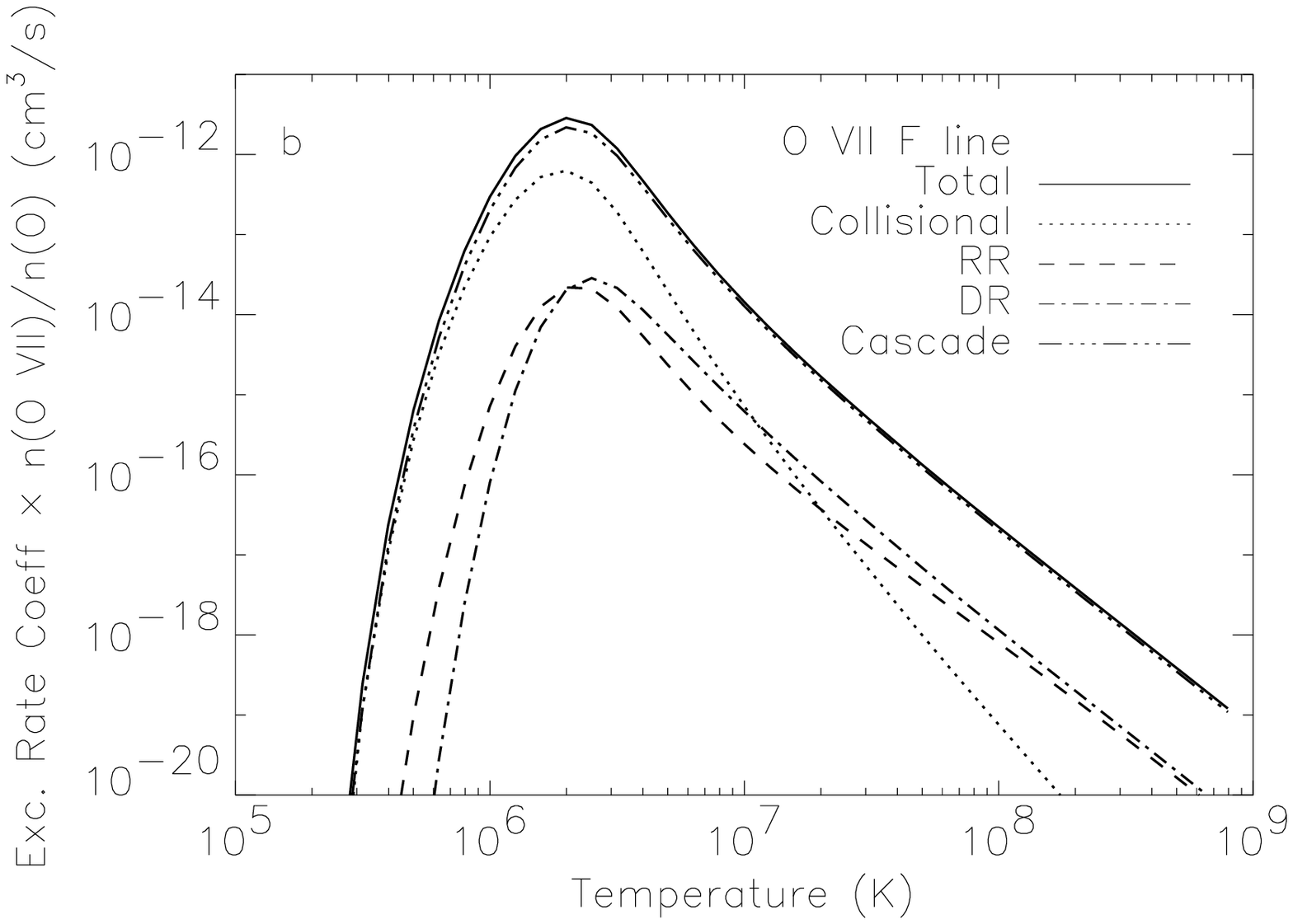}
\caption{(a) Excitation to the $1s2p\ {}^1\hbox{P}_1$\ level as a
function of electron temperature, with the ionization balance from
MM98 for \protect{\ion{O}{7}}\ included.  The total (effective) rate
coefficient is shown, as well as the contributions from direct
electronic collisions, radiative (RR) and dielectronic (DR)
transitions, and cascades.  The dominant process at all temperatures
is electron collisions. (b) Same, for excitation to the $1s2s\
{}^3\hbox{S}_1$\ level.  However, in this case electron collisions are
significant only at low temperatures.  From the temperature of peak
emissivity and beyond, cascades from higher levels dominate.
\label{fig:ExcRate}}
\end{figure*}

Of course, the Bethe limit is only applicable if the upper levels of
both emission lines are populated by direct excitation from the ground
state, and depopulated by radiative transitions.  In the He-like
\ion{O}{7}\ system, other processes are more important than direct
excitation at high temperatures for populating the $1s2s\
{}^3\hbox{S}_1$\ level (see Figure~\ref{fig:ExcRate}(b)). The
differences in the \ion{O}{7}\ G ratio among the codes at high
temperature appear to be largely due to differences in the
dielectronic recombination rates. Figure~\ref{fig:ExcRate} shows the
level population mechanisms for the $1s2p\ {}^1\hbox{P}_1$\ level
(which forms the resonance line) and the \ion{O}{7}\ $1s2s\
{}^3\hbox{S}_1$\ level (which decays to the ground state as the
forbidden line).  The total excitation rates for these levels are
equivalent to the line emissivities, since radiative decay to the
ground state is the primary de-excitation process at low electron
density for both levels. In Figure~\ref{fig:ExcRate}(a), we see that
direct excitation dominates all other mechanisms for populating the
$1s2p\ {}^1\hbox{P}_1$\ level.  However, Figure~\ref{fig:ExcRate}(b)
shows that for the $1s2s\ {}^3\hbox{S}_1$\ level, direct excitation is
important only at low temperatures.  At the peak emissivity, the main
excitation mechanism is cascades from higher levels.  These higher
levels are primarily populated by dielectronic recombination, which we
have checked by comparing the computed rates for all the processes.
As a result, the Bethe limit cannot be used as a check on the
high-temperature limits of this He-like line ratio.  This can also
explain the difference between the \apec+MM98 and pure-\apec
calculations shown in Figure~\ref{fig:LineRatio}(b).  As a result of
the different recombination rates ($\sim 30$\% variation), at
temperatures above $2\times10^6$\,K, the pure-\apec code has a
slightly larger \ion{O}{7} population and slightly smaller \ion{O}{8}
population.  Thus the resonance line R emissivity (which is
proportional to the \ion{O}{7} abundance) is increased, while the F+I
emission, due largely to dielectronic recombination from \ion{O}{8},
is reduced.  This leads overall to a small decrease in the G ratio.

\begin{figure*}
\includegraphics[totalheight=2.2in]{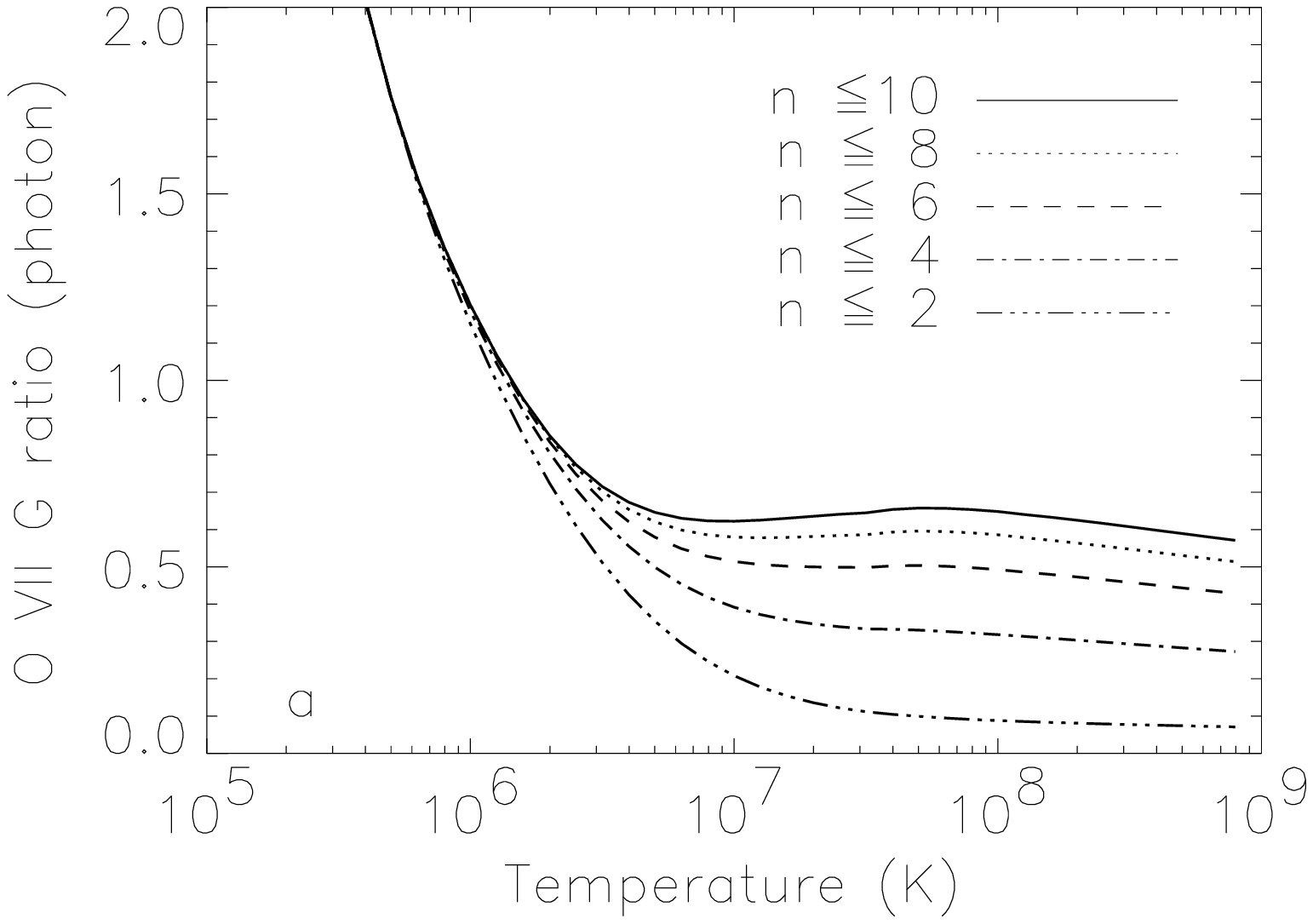}
\includegraphics[totalheight=2.2in]{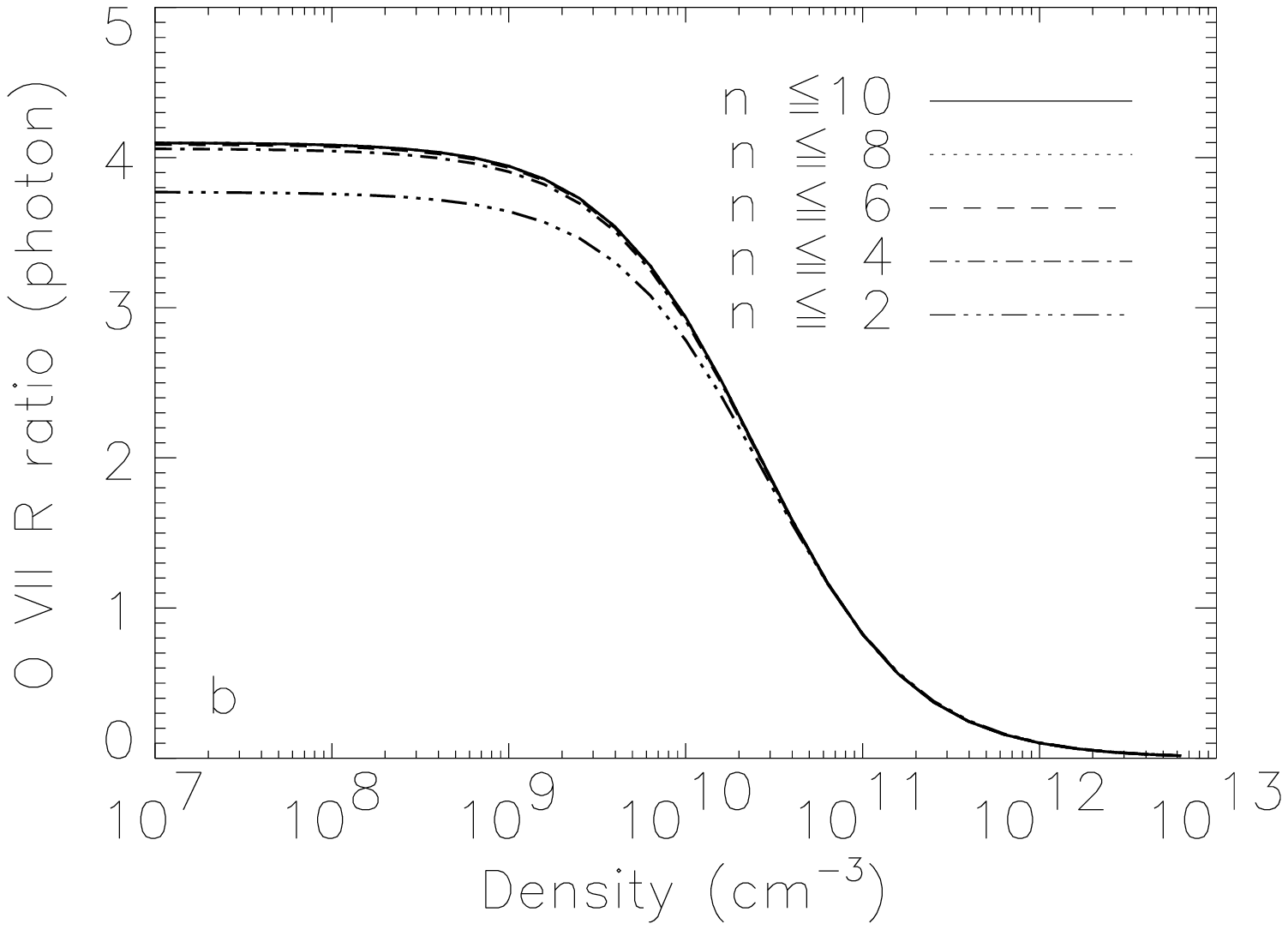}
\caption{(a) The \protect{\ion{O}{7}} $G$\ ratio at low electron
density as a function of temperature (using the MM98 ionization
balance), for five different atomic models which include levels with
the primary quantum number less than or equal to 2, 4, 6, 8, and 10.
By including levels up to n=10, the result has converged at high
temperatures to the accuracy of the underlying atomic data. (b) The
\protect{\ion{O}{7}}\ $R$\ ratio at $T = 10^6$\,K as a function of
electron density $n_e$, for the same five models.
\label{fig:ratio}}
\end{figure*}

For the H- and He-like isosequences of ions with significant cosmic
abundances (C, N, O, Ne, Mg, Al, Si, S, Ar, Ca, Fe, and Ni) \aped
includes atomic calculations for principal quantum numbers $n \le 5$.
We have tested the adequacy of this model by running \apec repeatedly
using only the $n \le 2$, $n \le 3$, and the $n \le 4$\ levels of
\ion{O}{7}, and find that, above the temperature of peak emissivity,
our results are not convergent.  We therefore obtained data for $n \le
10$ for \ion{O}{7}; the collisional excitation rate coefficients and
radiative rates are from HULLAC and the DR satellite line data are
from Safronova (2001, private communication).  The \ion{O}{7} G ratio
for each model, with $n \le 2, 4, 6, 8,$\ and 10 is shown in
Figure~\ref{fig:ratio}.  Clearly, the convergence including all $n \le
4$\ is inadequate.  By including all levels up to $n=10$\ we converge
to within the 10\%-20\% accuracy of the underlying atomic data.  In
the case of the $R=F/I$\ ratio, which is sensitive to the electron
density for $n_e \gtrsim 3\times10^9$\,cm$^{-3}$, fewer levels are
required to reach convergence at $T=10^6$\,K, as can be seen in
Figure~\ref{fig:ratio}(b).  At higher temperatures, such as
$T=6\times10^6$\,K, the spread among models is slightly larger, but
including only levels with $n \le 5$\ is still adequate.

\section{Conclusions}

This is the first in a sequence of papers on the new collection of
atomic data \aped and collisional plasma code \apec.  By separating
the code and data we can more easily test the convergence of our
models and compare atomic data from different sources.  We have shown
that, for \ion{O}{8}, using newer atomic data leads to substantially
different results in the high-temperature limit for the
Ly$\beta$/Ly$\alpha$\ ratio, which agrees with some recent EBIT
measurements. In addition, for He-like ions, high-$n$\ dielectronic
recombination from the H-like ion can significantly affect the
low-$n$\ line ratio G. Careful treatment of recombination to
individual levels and cascades is necessary for other diagnostic line
ratios; {\it e.g.} neon-like \ion{Fe}{17}\ (Liedahl 2000). The
methods described in this paper can be used to test the importance of
such detailed treatment. Furthermore, in cases where recombination and
cascades contribute significantly to the level population, the
accuracy of the ionization balance, not explicitly considered in this
paper, will contribute to the total uncertainty in the model line
ratio.

We thank Priyamvada Desai, Richard Edgar, Kate Kirby, Brendan
McLaughlin, and Ulyana Safronova for their assistance over the course
of creating \apec/\aped.  We gratefully acknowlege support from NASA
grant NAG5-3559 and the Chandra X-ray Center.  Work at LLNL was
performed under the auspices of the U. S. Department of Energy by
University of California Lawrence Livermore National Laboratory under
Contract No. W-7405-Eng-48.

\end{document}